\documentclass[prl,reprint,superscriptaddress,longbibliography,notitlepage,nofootinbib]{revtex4-1}
\usepackage{amsmath,amssymb}
\usepackage[euler]{textgreek}

\usepackage{psfrag}
\usepackage{graphicx}
\usepackage{color}
\usepackage[usenames,dvipsnames]{xcolor}
\usepackage{pifont}
\usepackage{fourier}
\usepackage{xr}
\usepackage{stmaryrd}
\usepackage{cancel}
\usepackage{mathrsfs}

\usepackage{tikz}
\usepackage{verbatim}

\usetikzlibrary{arrows,shapes,backgrounds}
\usepackage{tikzpagenodes} 
\usepackage{fancyhdr}

\usepackage[process=auto]{pstool}

\usepackage[normalem]{ulem}
\usepackage{euscript,array}

\newcommand{\bea}{\begin{eqnarray}}
\newcommand{\eea}{\end{eqnarray}}
\newcommand{\beq}{\begin{equation}}
\newcommand{\eeq}{\end{equation}}

\newcommand{\bega}{\begin{eqnarray}}
\newcommand{\eega}{\end{eqnarray}}

\newcommand{\be}{\begin{equation}}

\newcommand{\vf}{\varphi}

\newcommand{\g}[0]{\gamma}

\newcommand{\tr}[0]{\text{tr}}

\definecolor{redg}{rgb}{1,0,0}
\definecolor{blueg}{rgb}{0.22,0.33,0.64}
\definecolor{greeng}{rgb}{0,0.63,0.29}
\definecolor{orangeg}{rgb}{0.96,0.47,0.13}
\definecolor{figtext}{RGB}{102,102,102}

\DeclareMathAlphabet\mathbfcal{OMS}{cmsy}{b}{n}

\begin{document}

\title{Quantum Droplets in Curved Space}

\author{Antonino Flachi}
\email[]{flachi@phys-h.keio.ac.jp}
\affiliation{Department of Physics \& Research and Education Center for Natural Sciences, Keio University, 4-1-1 Hiyoshi, Kanagawa 223-8521, Japan}
\author{Takahiro Tanaka}
\email{t.tanaka@tap.scphys.kyoto-u.ac.jp}
\affiliation{Department of Physics \& Center for Gravitational Physics and Quantum Information, Yukawa Institute for Theoretical Physics,, Kyoto University, Kyoto 606-8502, Japan}
\begin{abstract}
This Letter investigates the formation of quantum droplets in curved spacetime, 
highlighting the significant influence of curvature on the formation and properties of these objects. While our computations encompass various dimensions, we primarily focus on two dimensions. Our findings reveal a novel class of curvature-driven quantum effects leading to the formation of quasistable liquid droplets, suggesting a feasible pathway for experimental observation, particularly in microgravity environments. 
\end{abstract}
\maketitle


\emph{Introduction.} 
The suggestion that a specific choice of the interaction strengths between atoms in binary mixtures could induce the formation of quantum liquid droplets has emerged as an important recent observation \cite{petrov:2015,astra:2016,Ferrier-Barbut:2018}, which has highlighted the essential role of quantum effects in explaining the existence and properties of matter under extreme (dilute) conditions and experimental observations have generated a certain amount of excitement within the cold-atom community and beyond (see, for example, Refs.~\cite{Nature.530.194,PhysRevLett.116.215301,Nature.539.259,Science.359.301,PhysRevLett.120.235301,PhysRevLett.120.160402,PhysRevLett.120.135301,Arazo:2021,Moss:2025xxq,PhysRevLett.121.195301,PhysRevLett.125.195302} or Refs.~\cite{Bottcher:2021,Chomaz:2023,Baroni:2024} for review). 
Here, we argue that quantum stabilized droplets provide direct access to a semiclassical \textit{quantum} regime, where geometrical effects play a pivotal role in altering the physics of these systems in a predictable and, in principle, observable way, allowing to test semiclassical quantum field phenomena in curved space.

In the case of a single-species, dilute interacting Bose gas, {Bogoliubov} long time ago \cite{Bogolubov.1947} calculated the ground state energy $\mathscr E$ of $\mathscr N$ bosons confined in a volume $\mathscr V$ with uniform density $n = \mathscr N/ \mathscr V$, interacting through a short-range $\delta$-function potential in the mean-field approximation,
\beq
\varepsilon = {g\over 2} n^2,
\label{eq:1}
\eeq
where $\varepsilon =\mathscr E / \mathscr V$ is the energy density and $g$ a coupling constant, indicating that, in dilute conditions and for fixed particle number, the atoms can exist only as a gas. The mean-field result (\ref{eq:1}) was extended to include quantum effects, a result due to {Lee, Huang, and Yang}  \cite{PhysRev.106.1135,PhysRev.106.1117,PhysRev.105.1119,PhysRev.112.1419}, 
\beq
{\varepsilon} =  {2 \pi a \over m} n^2 \left[1 +{128 \hbar\over 15 \sqrt{\pi}} \; \sqrt{n a^3}\right],
\label{eq:2}
\eeq
where $g= 4\pi a/m$ and $a$ the $s$-wave scattering length. 
The presence of this correction implies that the zero momentum part of the condensate containing most of the atoms undergoes zero-point quantum fluctuations, while a small fraction of atoms remain excited, a phenomenon known as \textit{quantum depletion}. In practice, quantum corrections do not change the nature of the atomic ensemble because the energy remains a monotonic function of the density. 
(See Refs.\cite{Pitae:2016,Toms_Schwinger} for reviews.)

The new observation of Ref.~\cite{petrov:2015} is that in a two species Bose gas adjusting the relative interaction strengths, in a way that makes same-species atoms weakly repel and different species atoms weakly attract, reduces the resulting leading-order contribution to the energy density and enhances, relatively, the quantum corrections. The outcome is remarkable as a minima in the energy emerges at a nonvanishing number density: A localized spherically symmetric configuration -- a droplet, stabilized by quantum effects -- forms. 

In this Letter, we consider a similar two species interacting field theory setup and include a generic curved background substrate. We shall demonstrate that the presence of curvature triggers a mechanism by which the stabilization (or destabilization) can be
driven or altered by geometrical effects. The role that geometry can play in the formation of droplets in particular, and in ultra-cold atoms, in general, is motivated by the possibility of using spherical trapping potential or a combination of optical traps to generate a curved background. In the context of quantum field theoretical approaches to Bose-Einstein condensation, curvature effects have been studied at least since Refs.~\cite{toms_bose_einstein,kirsten:1995}. See Refs.~\cite{tononi:2019, tononi:2023, Lundblad:2023, tononi:2024, tononi:2025} for more recent relevant work.

The qualitative effect of curvature on droplet properties can be understood, for instance, by dimensional arguments implying that additional contributions to the one-loop effective action, proportional to the Ricci scalar $R$ and the density, should occur. (In odd spacetime dimensions, logarithmic corrections are also expected). These curvature-dependent terms compete with chemical potential contributions, effectively shifting it by an amount proportional to $R$, indicating that curvature can significantly impact droplet stability, either stabilizing or destabilizing them depending on the sign and magnitude of $R$. These considerations are also supported by general properties of interacting quantum field theories in curved space (see Refs.\cite{Flachi:2011a, Flachi:2012, Flachi:2011, Flachi:2014, Flachi:2015, Flachi:2018}): The effective potential can be lifted by the presence of curvature, which in topologically simple cases, such as compact spaces, can be understood by analogy with finite-temperature effects. Collectively, these qualitative insights point toward a potentially substantial role for curvature and can be made more precise by explicit calculation.

\textit{Two species quantum gas in curved space.} 
In the following we consider a model describing a Bose-Bose mixture
\beq
{\mathscr S} =
\int dv \left\{\sum_p \left[ \Re\left(i \vf_p^* \dot{\vf}_p \right) 
-{\left|\nabla \vf_p \right|^2 \over 2m_p} 
\right] - \mathsf{U}
\right\},
\label{lag}
\eeq
where $\mathsf{U} = \sum_{pq}  {\gamma_{pq}} \left|\vf_p\right|^2 \left|\vf_q\right|^2/4$. 
The above model is assumed to live on a curved background spacetime, which we  assume to be static, 
defined by a metric tensor $g_{\mu\nu}$ ($g$ is the determinant and $\nabla$ is the scalar covariant derivative in curved space). Other than the assumption of staticity, the background is kept general. We have defined $dv = d^dx dt \sqrt{g}$ and $d$ is the spatial dimensionality. 
The couplings $\g_{pp} = 4 \pi a_{pp}/m_p$ and $\g_{12} = 4 \pi a_{12}(m_1 +m_2)/(m_1 m_2)$ are the intra- and interspecies coupling constants, respectively and $a_{pq}$ are the relevant scattering lengths. 
We set $\gamma_{12}=\gamma_{21}$ and adopt natural units where $\hbar =1$.  
Relation (\ref{lag}) represents a minimal (e.g., without nonminimal curvature couplings) extension of the model of Ref.~\cite{petrov:2015} to $d+1$ dimensional curved spacetime. 
Classically, in flat space the system is stable when $\mathsf{U}$ is bounded from below, which implies $\g_{11}>0$, $\g_{22}>0$ and $\g_{12}^2 < \g_{11} \g_{22}$, 
i.e., the quadratic form above is non-negative and the potential has a single minima at $\left|\vf_1\right| = \left|\vf_2\right| =0$. 
{The classical ground state is characterized by uniform condensate densities satisfying} 
$0 = \mu_p -{1\over 2} \sum_{q=1}^2 \g_{pq} \left|\bar\vf_q \right|^2$, for $p=1,2$, 
where $\mu_p$ represents the chemical potential associated to the species $p=1,2$. In curved space, we need (necessarily, in the case of a spatially varying background) to consider more general, inhomogeneous, configurations for the ground state; thus, we expand around a semiclassical space-dependent, static background $\bar \vf_p(x)$:
\beq
\varphi_p =  \bar \vf_p(x) + \delta{\vf}_p(t,x).
\eeq
The perturbations ${\delta \vf}_p(t,x)$ represent the quantum fluctuations that are integrated over in order to obtain the one-loop effective action. (See Refs.\cite{Toms_Schwinger,toms_bose_einstein,Toms:1995rr,kirsten:1996} for examples of application of the Schwinger effective action formalism to Bose-Einstein condensation and for other earlier references.) We introduce for convenience the notation 
\bea
{\bf \Phi}^{T} = \left( {\delta \vf}^{r}_1, {\delta \vf}^{i}_1, {\delta \vf}^{r}_2,  {\delta \vf}_2^{i}\right)
\eea
with each of the complex field components 
treated as an independent degree of freedom and corresponding to the real and imaginary parts of each of the perturbations (e.g., ${\delta \vf}^{r}_1$ is the real part of ${\delta \vf}_1$, etc.).  Implementing the expansion to second order and removing linear terms yields
\bea
{\mathscr S} \approx \bar{\mathscr S}  + 
\mathscr{S}_{,PQ} \left[\bar \vf_p  \right] \Phi_P \Phi_Q 
\label{secordpert}
\eea
where $\bar{\mathscr S} = {\mathscr S}\left[ \bar \vf_p \right]$; the approximation symbol signifies that the expansion ignores terms of order higher than second in the perturbations. We use the condensed notation of DeWitt \cite{DeWitt:1968}, where repeated indices are summed and spacetime variables, not indicated explicitly, are integrated over. The Lee-Huang-Yang correction can be calculated from the second term in (\ref{secordpert}) from the effective potential (see \cite{Toms_Schwinger} for an explicit calculation in the case of a single complex scalar field in $d=3$).

\textit{Quantum effects.} 
In the following, we want to isolate the effect of curvature in the quantum corrections and for simplicity we consider the case of identical species, e.g. 
$m_1=m_2=m$, and $\g_{11}=\g_{22}=g_s$ and $\g_{12}=\g_{12}=g_i$. If we write the background configuration as $\bar\varphi_p = \rho_p e^{i \theta_p}$ for $p=1,2$, we get for the action evaluated over the background
\bea
{\mathscr S}\left[ \bar \vf_p \right]
&=& 
 \int dv
\left\{ \sum_{p=1}^2\left[ -{1\over 2m} \left( \left(\nabla \rho_p \right)^2 + \rho_p^2 \left(\nabla \theta_p\right)^2\right)  
\right.\right.\nonumber \\
&& \left.\left. - {g_s \over 2}  \rho_p^4\right]
- {g_i }\left|\rho_1\right|^2\left|\rho_2\right|^2
\right\},\nonumber
\eea
and for the second-order correction 
\bea
\mathscr{S}_{,PQ} \left[{\vf_p} \right] \Phi_P \Phi_Q &=& 
\int 
dv\;
{\bf \Phi}^{\dagger} \mathsf{D}[{\vf_p}  ] \; {\bf \Phi}.
\eea
An explicit expression of the operator $\mathsf{D}[{\vf_p} ]$ can be found in formula (\ref{Dlong}) in the Supplemental Material.  
The one-loop effective action can then be written \cite{DeWitt:1968,Toms_Schwinger} as
\bea
\Gamma &=&
\bar{\mathscr{S}}
+ {1\over 2} \ln \det \left(\mathsf{D} [\bar\varphi_p ] \right).
\label{functionaldeterminant}
\eea
The quantum contributions that stabilize the droplet for specific choices of the curvature and of the couplings, come from the functional determinant, which depends on the ground state and has to be computed self-consistently by extremizing the full effective action as a function of $\bar\varphi_p$. Technically, in the present approach, it is the presence of cross-interaction terms that complicates the matrix structure of the operator leaving us, as we shall see, with the task of computing functional determinants of higher-order minimal operators. Here, since we selected a U$(1)$ symmetric potential and two identical species, we shall look for a ground state symmetric under the exchange of species ``flavor'' and assume that the two flavors have vanishing phases. 
Also, since we look for ground states that are not rapidly varying functions in space, spatial derivatives emerging from higher order terms as perturbations. 

Some remarks on how the computation can be carried out, in general, and how derivatives can be treated systematically in the present approach are illustrated in the Supplemental Material.  
Our choice of ground state yields
\bea
{1\over 2} \ln \det \left(\mathsf{D}\left[\rho\right] \right)
= 
{1\over 2} \ln \det \mathsf{D}_+
+
{1\over 2} \ln \det \mathsf{D}_-
\label{detpmaux1}
\eea
where 
$$
\mathsf{D}_\pm = 4m^2 {\partial^2 \over \partial t^2} 
- \Delta^2   + 4m \mathsf{X}_\pm \Delta + 4 m^2 \mathsf{S}_\pm
$$ 
with
\bea
\mathsf{X}_\pm &=& 
- 2 g_s \rho^2 -2 g_i \rho^2 \delta_{+k},\\
\mathsf{S}_\pm &=& 
\left((g_i - 3 g_s)  - 4 g_i \delta_{+k} \right)(g_i + g_s) \rho^4,
\eea
with $k=\pm$ and $\delta_{+k}$ being a Kronecker delta. {$\Delta$ is the Laplace-Beltrami second-order operator. We follow notation and conventions as in Ref.~\cite{Parker:2009uva}.} 
The determinants in (\ref{detpmaux1}) can be calculated using zeta-function regularization that has the advantage to allow for incorporating effects of curved space and of inhomogeneities systematically \cite{Kirsten:2004,Elizalde:1994,Parker:2009uva,Avramidi:2015}. The determinants can be expressed, upon Euclideanization $t \to -i \tau$ with $\tau \in \mathbf{S}^1\left(\beta\right)$ with $\beta$ being the inverse temperature, in the form ($k=\pm$)
\bea
\ln \det \mathsf{D}_k
&=& - \zeta\left[\mathsf{D}_k\right]'(0)-  \zeta\left[\mathsf{D}_k\right](0) \ln \xi^{-2},
\label{zetafull}
\eea
where $\xi$ is a renormalization scale, which in the present case has mass dimension $-2$. The generalized zeta function in (\ref{zetafull}) can be expressed via the Mellin integral representation 
\bea
\zeta\left[\mathsf{D}_\pm\right](s) \equiv \tr\; \mathsf{D}_\pm^{-s} = {1\over \Gamma(s)}\int_0^\infty {dz \over z^{1-s}} \theta\left(4 \pi z/\tilde\beta^2 \right) \tr\; e^{-z \mathscr{F}_\pm\left[ \Delta\right]},
\label{funtr}
\eea
where we have introduced
$\mathscr{F}_\pm\left[ \Delta\right]= {\Delta^2} - 4 m X_\pm {\Delta} 
+ 4m^2 S_\pm^2$
and the Jacobi elliptic 
function
$\theta\left(x\right) = \sum_{n\in \mathbf{Z}} e^{- \pi x n^2}$, 
with $\tilde\beta^2 = \beta^2/(4m^2)$, $x=4 \pi z/\tilde\beta^2$. 
In order to separate the finite-temperature from the zero-temperature part, we may perform a modular transformation $\theta\left(u\right) = \sqrt{1/ u}\;\theta(1/u)$ and use a series representation of the theta function, leading to
\bea
\theta\left(4 \pi z/\beta^2\right)
=
\sqrt{\tilde\beta^2\over 4 \pi z} \left(1 + 2 \sum_{n=1}^\infty \exp\left(- {\tilde\beta^2 n^2\over 4 z}\right)\right).
\label{thetanull}
\eea
The procedure we follow consists in expanding the integrand in (\ref{funtr}) in terms of the ground state; this can be done by using the asymptotic expansion (we rely here on Ref.~\cite{Barvinsky:2021}, whose results are used here in a slightly modified form by exponentiation of the trace):
\bea
\tr e^{-z \mathscr{F}_\pm\left[ \Delta\right]} = \sum_{q=0}^\infty z^{{q\over 2}-{d\over 4}} \int {d^dx \over (4\pi)^{d/2}}
{g^{1/2}} {\Gamma(d/4) \over 2 \Gamma(d/2)}  
e^{-\sqrt{z} \mathsf{T}_\pm}
\check{\mathsf{E}}_{2q}^{(\pm)},~~~ 
\label{exptrace}
\eea
where $\mathsf{T}_\pm = - {\Gamma(d/4+1/2) \over d \Gamma(d/4)} \left\{ \mathsf{D} +{d\over 3}R \right\}$ and $\mathsf{D} = -4d m X_{\pm}$. The first two coefficients\footnote{Details on the asymptotic expansion of the functional trace  involving fourth-order minimal operators and a formula for $\check{\mathsf{E}}_{4}$ are discussed for completeness in the Supplemental Material.} are $\check{\mathsf{E}}_{0}^{(\pm)}=1$ and $\check{\mathsf{E}}_{2}^{(\pm)}=0$. 
{In the following, we shall perform a truncation and ignore terms of mass dimension greater than 6, i.e. $k\geq 3$. This level of approximation implies that the background geometry cannot be highly curved, in the sense of $O(R^3)$, or highly inhomogeneous, in the sense of $O(\nabla \nabla R^2)$. Also, combinations between curvature and ground state ignored within the present approximation consist of $\left(\nabla \rho^2\right)^2$, $\Delta^2 \rho^2$, $\Delta R  \rho^2$, $R^4 \rho^2$ and higher powers of the ground state. Physically, this means that our approach cannot be applied to highly curved or strongly inhomogeneous backgrounds and reveal whether there exists a ground state that is rapidly varying in space; while this seems unlikely, this is, in fact, a limitation of the present approach that we accept in view of the fact that droplet profiles are indeed not rapidly varying functions in space by definition. We stress that this argument can be made rigorous by computing the next-order corrections $k=3$ that could allow one to improve the approximation perturbatively. However, including highly curved and inhomogeneous backgrounds requires a more extensive use of numerical analysis. It helps intuition to recall that the idea behind this work is that even a small curvature can shift, in principle, the effective potential: If the droplet is obtained close to a critical configuration, then even a small curvature can further stabilize or destabilize the droplet.}

Substituting (\ref{exptrace}) in (\ref{funtr}) and integrating over $z$, we find
\bea
\zeta_{\mathsf{D}_\pm}(s) = \int dv\; 
{ {\beta / (2m)}\over  \left({4 \pi}\right)^{(1+d)/2}}
 {\Gamma\left(d/4\right) \over \Gamma\left(d/2\right)}
\sum_{q=0}^\infty 
\left[
\mathscr{Z}^{(\pm)}_q (s) + \mathscr{T}^{(\pm)}_q (s)
\right]. \label{deltaZ}
\eea
The contributions $\mathscr{Z}^{(\pm)}_q (s)$ and $\mathscr{T}^{(\pm)}_q (s)$ can be calculated exactly (See Supplemental Material for details), yielding for the zero-temperature term
\bea
\mathscr{Z}^{(\pm)}_m (s) &=&  
{\Gamma\left(2s + q -d/2 -1\right) \over \Gamma\left(s\right)} \left|\mathsf{T}_\pm\right|^{1-q+d/2-2s}\check{\mathsf{E}}_{2q}.
\eea
The finite-temperature contribution $\mathscr{T}^{(\pm)}_m (s)$ can also be computed exactly, and its explicit form is presented in Supplemental Material for completeness, as it is lengthy and not central to the present discussion. The above results already convey a certain amount of information. First of all, taking the $s\to 0$ limit reveals that $\zeta_{\mathsf{D}_\pm}(0)=0$ only for $d$ odd (e.g., $d=1, 3$): Logarithmic contributions appear only for $d$ even (e.g., $d=2$). This general property is in agreement for $d=1, 2$, and $3$ with Refs.~\cite{petrov:2015,astra:2016}: Within zeta-regularization, as usual in spaces without boundaries, the occurrence of logarithms in the one-loop corrections is understood as related to the dimensionality. The thermal part coming from $\mathscr{T}^{(\pm)}_m (s)$ does not contribute to $\zeta_{\mathsf{D}_\pm}(0)$ as expected from general arguments (as thermal corrections give rise to finite contributions). Here, we are interested in the $T=0$ case and ignore the thermal contribution in what follows. 
Using (\ref{functionaldeterminant}), (\ref{detpmaux1}), (\ref{zetafull}), and (\ref{deltaZ}), we arrive at the one-loop effective action:
\bea
\Gamma &=&
\bar{\mathscr{S}}
- {1\over 2}  \sum_\pm
\left[
\zeta'_{\mathsf{D}_\pm}(0)
+
\zeta_{\mathsf{D}_\pm}(0)
\ln \xi^{-2}
\right]
\equiv 
\bar{\mathscr{S}}
+
\tilde \Gamma
\label{fullGamma}
\eea
where the latter equality defines $\tilde \Gamma$ and is introduced for brevity of notation. For example, for $d=3$ the previous expression reduces to
\bea
\tilde \Gamma &\approx&
-{\beta \over m}  {\Gamma(3/4) \over 240 \pi^2}  
\int dv
\sum_{k=\pm} 
\left[
4 \left| \mathsf{T}_k \right|^{5/2} +
15 \left|\mathsf{T}_k\right|^{1/2} \check{\mathsf{E}}_{4}^{(k)} 
\right],~~~
\label{d3}
\eea
where $\approx$ signifies that we have truncated the expansion to $q=2$ (i.e., we have excluded terms of mass dimension greater than $6$).

\textit{Curvature-driven droplets on surfaces.} 
Here, we are primarily interested in droplets forming on surfaces, $d=2$:
\bea
 \Gamma &=& - {\beta\over 128 \pi m}  
\int dv
\sum_{k=\pm} \left[
\left| \mathsf{T}_k \right|^2 \left(\gamma_e -3 + \ln\left(\left| \mathsf{T}_k \right|^2 \xi^2\right) \right)  \right.
\nonumber\\
&-&\left.
\left( {R^2 \over 18} + {S_k} + {7\over 6} RX_k\right)\left(\gamma_e + \ln\left(\left| \mathsf{T}_k \right|^2 \xi^2\right)\right)
\right]. 
\label{d2}
\eea
Notice that taking the limit of vanishing curvature and appropriately choosing the renormalization scale $\xi$ allows one to recast the previous expression into the following form:
\bea
\tilde \Gamma &=& - {\beta \varpi\over 2 \pi m}  
\int dv n^2 \left[
\ln \left(n^2/\hat{n}_0^2\right) - 1 
\right], \label{d2flat}
\eea
where we have defined $n = \rho^2$, rescaled $\xi$ in terms of $\hat{n}_0$, and defined $\varpi = g_s^2 +\pi g_s \delta g /16 + (1+\pi /32) \delta g^2$, where we have defined $g_s + g_i = \delta g$. {In the present case, it is possible to rescale $\xi$ in terms of the equilibrium density $n_0$ and the background field, allowing one to relate our renormalization scale to the scattering lengths. (To be precise, in (\ref{d2flat}) we have rescaled $\xi$ in terms of $\hat{n}_0$. It is easy to see that one can write (\ref{d2flat}) precisely as formula (6) in Ref.\cite{astra:2016}, by rescaling the ground state and $\hat n_0$. In this way, the apparent renormalization scale dependence ambiguity is eliminated. This conclusion is robust since the renormalization scale dependence does not reappear including higher-order terms.)}

\begin{figure}[t]
\centering
\includegraphics[width=0.99\columnwidth]{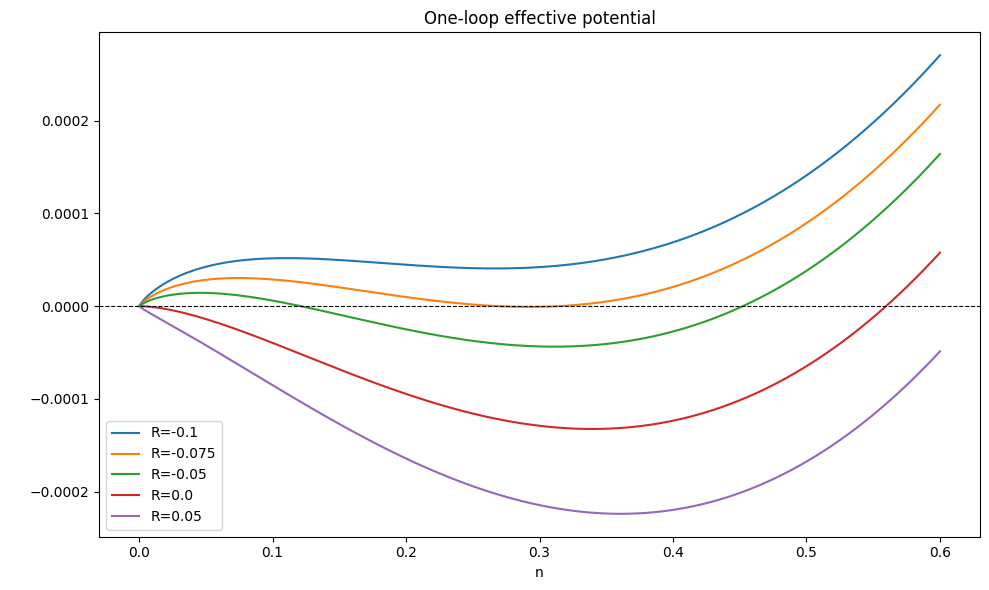}
\caption{\label{fig1} One-loop effective potential dependence from curvature. We have chosen for illustration $g_s=0.150$, $g_i = -0.145$ and $\xi=1$. Everything is measured in units of $m$. The value of the renormalization constant $\xi$ can be rescaled, resulting in a rescaled location of the minima. The plot illustrates the change of the potential from a $R>0$ nonvanishing minima to the flat space droplet minima through the transition happening for $R<0$.}
\label{fig-01}
\end{figure}

We can inspect the role of spatial curvature quite simply by looking at the one-loop effective potential that we can directly read off from (\ref{fullGamma}) and (\ref{d2}). If we choose the couplings $g_s>0$ and $g_i<0$ with $\delta g = g_s+g_i > 0$ and small to guarantee the existence of a droplet in flat space (in other words that the effective potential has a minima at a nonvanishing $\rho$ for $R=0$), then we can inspect how the minima moves when $R\neq 0$. Figure~\ref{fig-01} shows the changes in the effective potential for the given choice of parameters: The $R=0$ curve shifts downward (upward) for positive (negative) small $R$. For values of the curvature small but sufficiently negative, the nonzero minima (i.e., the droplet) disappear, through a first-order quantum phase transition. 
This behavior is due to the $R\rho^2$ correction in the potential, which is peculiar to curved space. Computing the derivative of the potential $U'_0 = \lim_{n \to 0} U'(n)$ keeping curvature terms up to O$(R \log (\left|R\right|))$ gives $U'_0 = (g_s + \delta g) R \left(
(16 + 7 \pi) \left(\gamma_e + \log \left(R^2 /(9\pi)\right)- 8 (4 + \pi)\right)/(384 \pi^2)\right)$ that is positive (negative) for  
negative (positive) $R$. We should remark that this behavior can be inverted for a configuration where the difference between $g_i$ and $g_s$ is fine-tuned, whereby positive curvature pushes the potential upward. For $\left|g_i \right|$ larger than $g_s$, the effect of the curvature may be expected to produce a small potential barrier between the minima at $n=0$ and larger values of the density, beyond which repulsion dominates. This raises the question as to whether it is possible to modify the interaction potential in a way that positive curvature leads to a destabilization of the droplet. In this case, the stabilization of a droplet configuration is not obviously guaranteed at least within the simple model analyzed here.
\begin{figure}[h]
\centering
\includegraphics[width=.99\columnwidth]{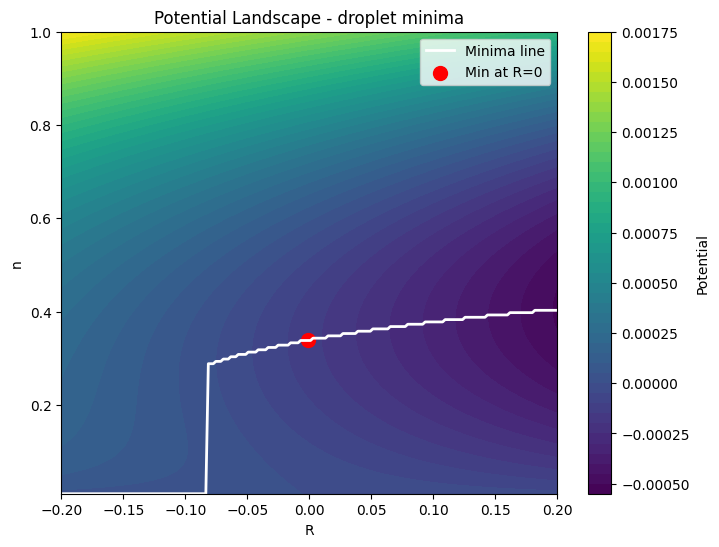}
\caption{\label{fig2} The figure illustrates the potential landscape, and the position of the droplet minima is indicated by the line. The dot represent the position of the flat space droplet minima, and the curve indicates how this minima moves when the curvature changes. The parameters have been set as in Figure~\ref{fig-01}. The transition point where the droplets are destabilized and the minima move to zero happens for a negative value of the curvature.}
\end{figure}

{We should notice that if the background geometry is inhomogeneous, then higher derivatives of the ground state may become important and could, in principle, be the drivers of dynamical instabilities. While we are not interested in this possibility, it is a problem that is worth mentioning in the context of a curved space formulation.}

{The results of this Letter refer to the case of a symmetric mixture. It may be interesting to extend the results of this work to include a small perturbation due to asymmetry in the mixture. The combined effect is difficult to anticipate, however, if the asymmetry is small one can expect that the droplet can receive larger contributions from
the species in excess, leading to a slight change in the profile of the droplet (if the asymmetry is
a perturbation). A slight change in the total energy of the droplet could
be viewed as an unbalancing between the surface tension and the energy associated with the
volume or area of the droplet. Since curvature can shift the energy, curvature effects can, in principle, offset changes due to small asymmetries, but a more detailed analysis is necessary to capture these aspects precisely.}

\begin{figure}[h]
\centering
\includegraphics[width=1.\columnwidth]{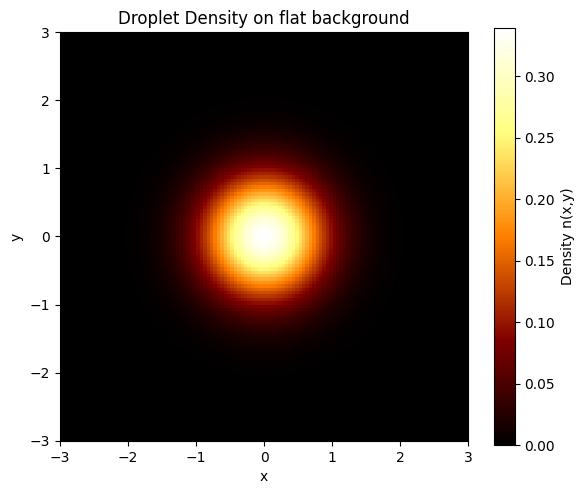}
\caption{Flat space, droplet density. The values of the coupling have been set for this and the other figures to $g_s=0.150$ and $g_i=-0.145$ and $\xi=1$. The peak value of the droplet density corresponds to the density at the minima of the effective potential, which in this case is $0.3395$.}
\label{fig-03-1}
\end{figure}

\begin{figure}[h]
\centering
\includegraphics[width=1.\columnwidth]{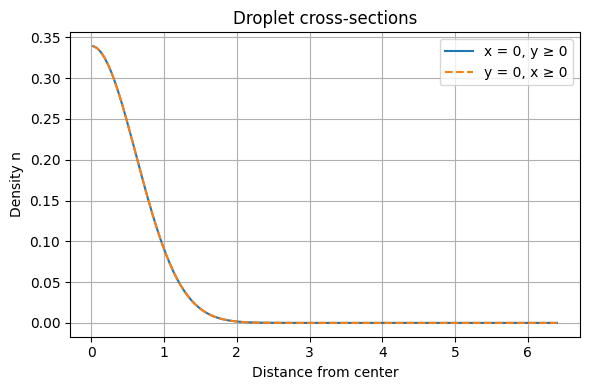}
\caption{Flat space, droplet density cross section along $x=0$ and $y=0$.}
\label{fig-03-2}
\end{figure}

\begin{figure}[h]
\centering
\includegraphics[width=1.\columnwidth]{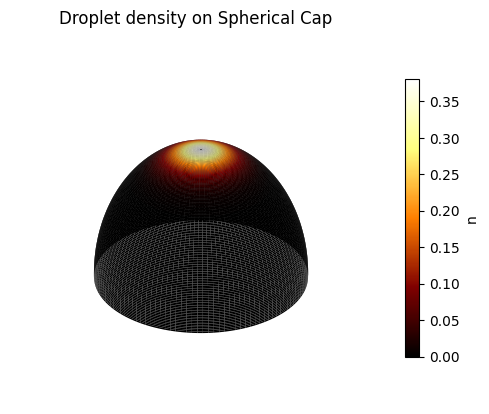}
\caption{Droplet density on spherical cap. Parameters have been set as in the other figures and the Ricci scalar is set equal to $2/3^2$. The peak value of the droplet density is in this case $0.3815$. 
}
\label{fig-04-1}
\end{figure}

\begin{figure}[h]
\centering
\includegraphics[width=1.\columnwidth]{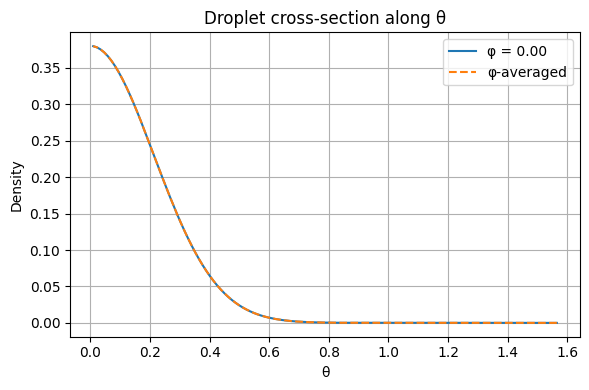}
\caption{Droplet density cross sections along $\theta$ (for $\varphi=0$ and averaged over $\varphi$).}
\label{fig-04-2}
\end{figure}

\begin{figure}[h]
\centering
\includegraphics[width=1.\columnwidth]{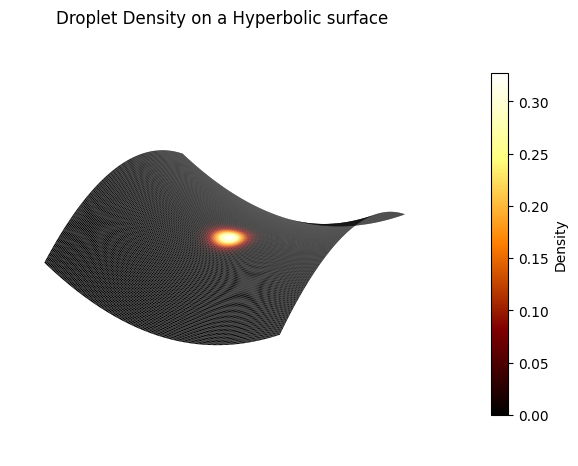}
\caption{Droplet density on negatively curved surface. Values of the coupling and other physical parameters are set as in the other figures. The Ricci is set to $R=-2/5^2$, which guarantees that there is a minima in the potential for nonvanishing density. The peak value of the potential is $0.3292$.}
\label{fig-05-1}
\end{figure}

\begin{figure}[h]
\centering
\includegraphics[width=1.\columnwidth]{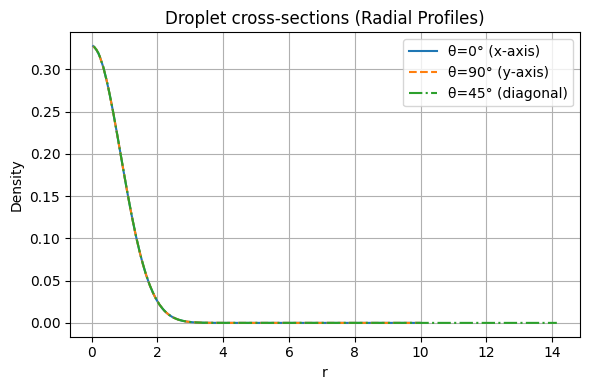}
\caption{Droplet density cross sections (radial profiles along $\theta$ (for $\varphi=0, \pi/4, \pi/2$).}
\label{fig-05-2}
\end{figure}

For completeness, we explicitly construct numerically some examples of static droplets solutions on weakly curved backgrounds, where the Ricci scalar is taken to be small, i.e., the curvature radius in the region where the droplet forms is large. The existence of such solutions is guaranteed by the form of the potential (i.e., the existence of a nonvanishing minima at nonzero density) and particle conservation. Here, we obtain numerical solutions using a pseudospectral solver that integrates Gauss-Legendre quadrature with fast Fourier transforms, with time evolution handled via the Strang-Yoshida splitting method \cite{Strang:1968,Yoshida:1990,McLachlan:2002,Hairer:2002}. 
Illustrative solutions are shown in Figs.~\ref{fig-03-1} and \ref{fig-03-2} for the case of droplets in flat space, in Figs.~\ref{fig-04-1} and \ref{fig-04-2} for the case of a weakly curved spherical cap and in Figs.~\ref{fig-05-1} and \ref{fig-05-2} for the case of a negatively curved sector of hyperbolic space (i.e., a 2D saddle).

\textit{Physical considerations.} 
We have explored how spatial curvature affects droplet formation in two-species interacting quantum field theories. Our findings reveal that curvature plays a significant role: It can shift the minima associated with droplet configurations and destabilize them for sufficiently negative curvature. This is noteworthy for two reasons. First, the destabilization of a droplet configuration in flat space points to the occurrence of a \emph{curvature-induced} first-order quantum phase transition. Second, even a slight positive curvature can trigger droplet formation in (near-critical) parameter regimes where, in flat space, droplets would not typically emerge. These results suggest that the dependence of quantum droplets properties on local spatial curvature could be observed in microgravity environments in space \cite{Alonso:2022}. It is interesting to try to generalize the picture to properly include gravitational effects and it is tempting to think about the possibility of a quantum-stabilized droplet as a device to detect gravitational waves. A more down-to-earth observation is that it is natural to expect that a similar phenomena should occur in models of bubble nucleation more complex than what has been so far done (i.e., with several species and unbalanced interactions).

A final pertinent question is whether observing such curvature-induced quantum phase transitions is feasible with current experimental techniques. 
{To have an idea of the typical scales involved, let us assume a typical droplet size of the order of 10 $\mu\text{m}$ 
and thicknesses of the order of a micron
and setting the mass scale in our model to be of the order of the inverse thermal de Broglie wavelength $\lambda$ at 100 nK, that is,
$\lambda \approx {h}/({\sqrt{2 m k_B T}}) \approx \text{O}(1 - 10 \mu\text{m})$, with $h$ and $k_B$ being the Planck and Boltzmann constants. Taking $0<m g_s \sim \text{O}(10^{-1})$ and $0>m g_i$ slightly smaller in absolute value suggests that a local curvature radius of $\text{O}(1 - 10 ) \mu\text{m}$, corresponding to a Ricci curvature of about $10^{-2} - 10^{-1} \mu\text{m}^{-2}$, should be within the realm of detectability. Practically, creating a curved optical trap with a radius of curvature around $10 \mu\text{m}$ is challenging but feasible utilizing optical shaping with spatial light modulators, digital micro-mirror devices, holography, or high-numerical aperture (high-NA) focusing. (See, for example, Refs.~\cite{Gauthier:2016,Gauthier:2021,Bowman:2015}.) Interestingly, Ref.~\cite{Viermann} realizes a 2D regime by using a detuned optical lattice that provides strong axial confinement leading to a thickness of $0.4 \mu\text{m}$, ensuring the system behaves as effectively 2D, with negligible out-of-plane excitations. A curvature is implemented through the BEC's density profile, mapped to curved metrics (hyperbolic or spherical). In this experiment the key length scale is the Thomas-Fermi radius of the condensate cloud, which sets the effective radius of curvature. While not suggesting any specific experiment, the value for the hyperbolic geometry quoted in Ref.~\cite{Viermann} is a Ricci scalar of $R\approx -3.2 \times 10^{-3} \mu \text{m}^{-2}$, which suggests, from a purely geometrical viewpoint, that the effect discussed here could potentially be measurable within the framework of Ref.~\cite{Viermann}.}

\textit{Acknowledgments.} 
The support of the Japanese Society for the Promotion of Science (Grant-in-Aid for Scientific Research, KAKENHI, Grants No. 21K03540, No. JP23H00110,  No. 24H00963. and No. 24H01809) is gratefully acknowledged. Thanks are due Matthew Edmonds for many discussions on various aspects of quantum droplets.\\

\newpage

\appendix
\thispagestyle{empty}
\begin{widetext}
\section{Supplemental Material -- Quantum droplets in curved space}

\subsection{Second order expansions}

The expansions are algebraically trivial, albeit a bit tedious. By substituting $\varphi_p =  \bar \vf_p(x) + \delta{\vf}_p(t,x)$ and subsequently expanding up to second order (i.e., eliminating terms or order 3 or higher in the perturbation and terms of linear order), we can obtain the operator $\mathsf{D}$ for a general time-independent complex ground state $\bar \vf_p(x)$:
\bea
\mathsf{D} 
= {\tiny \begin{pmatrix}
-{\Delta \over 2m_1} - \mu_1 - {g_s} \left( \left|\bar\varphi_1\right|^2 +  2 \Re \left[\bar\varphi_1\right]^2
\right)  - g_i \left|\bar\varphi_2\right|^2& i {\partial \over \partial t} 
- 2{g_s} \Re\left[\bar\varphi_1\right]\Im\left[\bar\varphi_1\right]   & -2{g_i} \Re \left[\bar\varphi_1\right]\Re \left[\bar\varphi_2\right]& - 2 {g_i} \Re \left[\bar\varphi_1\right]\Im \left[\bar\varphi_2\right]
\\
-i {\partial \over \partial t} - 2{g_s} \Re\left[\bar\varphi_1\right]\Im\left[\bar\varphi_1\right]   & -{\Delta \over 2m_1} - \mu_1 - {g_s} \left( \left|\bar\varphi_1\right|^2 +  2 \Im \left[\bar\varphi_1\right]^2
\right)  -g_i \left|\bar\varphi_2\right|^2& -2 g_i \Im \left[\bar\varphi_1\right]\Re \left[\bar\varphi_2\right]& - 2 {g_i} \Im \left[\bar\varphi_1\right]\Im \left[\bar\varphi_2\right]
\\
-2{g_i} \Re \left[\bar\varphi_1\right]\Re \left[\bar\varphi_2\right] &-2 g_i \Im \left[\bar\varphi_1\right]\Re \left[\bar\varphi_2\right] &-{\Delta \over 2m_2} - \mu_2 - {g_s} \left( \left|\bar\varphi_2\right|^2 +  2 \Re \left[\bar\varphi_2\right]^2
\right)   - g_i \left|\bar\varphi_1\right|^2& i {\partial \over \partial t} - 2{g_s} \Re\left[\bar\varphi_2\right]\Im\left[\bar\varphi_2\right]    
\\
- 2 {g_i} \Re \left[\bar\varphi_1\right]\Im \left[\bar\varphi_2\right]& -2 g_i \Im \left[\bar\varphi_1\right]\Im \left[\bar\varphi_2\right] &-i {\partial \over \partial t} - 2{g_s} \Re\left[\bar\varphi_2\right]\Im\left[\bar\varphi_2\right]   & -{\Delta \over 2m_2} - \mu_2 - {g_s} \left( \left|\bar\varphi_2\right|^2 +  2 \Im \left[\bar\varphi_2\right]^2
\right)  - g_i \left|\bar\varphi_1\right|^2
\end{pmatrix} }
\label{Dlong}~~~
\eea
From the above expression we can observe that the in the functional determinant we have to deal, modulo simplifications stemming from specific assumptions on the ground state, with high order operators. In the present work, we have assumed that the background geometry is sufficiently smooth and therefore we can treat derivatives of the ground state as perturbations. This choices, together with the assumption that the modulus $\left|\bar\varphi_1\right| = \left|\bar\varphi_2\right| = \rho$, allows to break the determinant into the product of two determinants of fourth order minimal operators. A more general assumption on the ground state would be necessary if one assumes that underlying geometry to be singular, or in the presence of defects, torsion, or to have  large spatial inhomogeneities, i.e. $\nabla \mathcal R \gg \mathcal R$ with $\mathcal R$ indicating a typical curvature scale.

\subsection{Heat-kernel asymptotics of fourth order minimal operators}

Zeta-function regularization allows us to express the one-loop effective action in terms of the functional trace $\tr e^{-z \mathscr{F}_\pm\left[ \Delta\right]}$ of a fourth-order minimal operator, which in our case is given by 
\bea
\mathscr{F}_\pm\left[ \Delta\right]= {\Delta^2} - 4 m X_\pm {\Delta} 
+ 4m^2 S_\pm^2.
\label{opF}
\eea
Our approach to compute the trace is based on the use of a mixed-derivative expansion, the heat-kernel expansion, which allows us to get a truncated expression in terms of the background field. This is a well established technique in quantum field theory in curved space and it has been used in similar computations involving second order operators (see for example \cite{Gilkey:1984,Vassilevich:2003xt,Kirsten:20043,Parker:2009uva4,Avramidi:20155} for more or less mathematically oriented references), the analogous expansion for a higher order operator is less explored (see, however, Refs.~\cite{Gilkey:1980,Fegan:1985,Branson:1985,Barvinsky:1985an,Gusynin:1988zt,Gusynin:1990ek} for some examples of earlier explorations and the more recent papers Refs.~\cite{Barvinsky:2021ijq,Barvinsky:2017mal} for a longer list of references. Here we follow Ref.~\cite{Barvinsky:2021ijq}. Notice however that the results needed in our calculations only require a simpler version of the more general results of Ref.~\cite{Barvinsky:2021ijq}. 

Consider a generic minimal fourth-order operator,
\bea
\mathsf{F}\left(\nabla \right)
= \Delta^2 + \mathsf{D}^{ab} \nabla_a \nabla_b + \mathsf{P}.
\eea
Comparing the above expression with (\ref{opF}) we see that for our case
\bea
\mathsf{D}^{ab} = - 4 m X_\pm g^{ab} ;~~~\mathsf{P} = 4m^2 S_\pm^2.
\eea
The heat-kernel diagonal, i.e., the coincidence limit, is expressed as
\bea
\mathsf{K}\left(z|x,x\right) = \tr e^{-\tau \mathsf{F}}
=
{g^{1/2}(x)  \over (4\pi)^{d/2}} {\Gamma(d/4) \over 2 \Gamma(d/2)} \sum_{m=0}^\infty z^{m-d/2 \over M} \mathsf{E}_{2m}(x),
\label{hkexp}
\eea
with $\mathsf{E}_{2m}(x)$ representing the generalization for the present case of the Seeley-Gilkey coefficients. The explicit expression of this coefficients has been computed in Ref.~\cite{Barvinsky:2021ijq}:
\bea
\mathsf{E}_{0}(x) &=& \hat{\mathsf{1}},\; ~~~\mathsf{E}_{2}(x) = 
{2 \Gamma(d/4+1/2) \over \Gamma(d/4)}
\left\{{1\over 2d}\mathsf{D} +{\hat{\mathsf{1}}\over 6}R\right\} \\
\mathsf{E}_{4}(x) &=& 
\left\{
{d-2\over 2}\left(
{\hat{\mathsf{1}}\over 90} R_{abcd}^2 - {\hat{\mathsf{1}}\over 90} R_{ab}^2
+{\hat{\mathsf{1}}\over 36} R^2 + {\hat{\mathsf{1}}\over 6} \mathcal{R}_{ab}^2
+{\hat{\mathsf{1}}\over 15} \Delta R\right)
+{{1}\over 12} \mathsf{D}^{ab} R_{ab}
+{{1}\over 12} \mathsf{D} R \right.\\ \nonumber
&&\left.
{1\over 2(d+2)}\left(
{1\over 2}\mathsf{D}_{ab}\mathsf{D}^{ab} + {1\over 4}\mathsf{D}^2 + {2\over 3}(d+1) \nabla_a\nabla_b\mathsf{D}^{ab} + {1\over 6}(d+4)\Delta \mathsf{D} 
\right) -\mathsf{P}
\right\} 
\eea
where $\mathsf{D} = \mathsf{D}^a_a$ and $\hat{\mathsf{1}}$ is the identity. 

In the computations presented in the main text, we have used the following form heat-kernel expansion
\bea
\tr e^{-z \mathscr{F}_\pm\left[ \Delta\right]} = \sum_{q=0}^\infty z^{q/2-d/4} \int {d^dx \over (4\pi)^{d/2}}
{g^{1/2}} {\Gamma(d/4) \over 2 \Gamma(d/2)}  
e^{-\sqrt{z} \mathsf{T}_\pm}
\hat{\mathsf{E}}_{2q}, 
\label{exphk}
\eea
which contains an exponential term; $\mathsf{T}_\pm = - {\Gamma(d/4+1/2) \over d \Gamma(d/4)} \left\{ \mathsf{D} +{d\over 3}R \hat{\mathsf{1}}\right\}$. The above expression differs slightly from that used in Ref.~\cite{Barvinsky:2021ijq} due to the exponential that we have inserted in the expansion. For second order operators such exponential insertions are used to re-sum powers of the background field in the heat-kernel series (see Refs.~\cite{Parker:1984dj,Jack:1985mw} and Refs.~\cite{Fecit:2025kqb,Franchino-Vinas:2023wea,Ferreiro:2020uno} for a few sample applications). In our computation the exponential can be viewed as nothing but a redefinition of the coefficients. In fact, within our scheme, and particularly for $d=2$ spatial dimensions, this redefinition allows to express remove the second term in the heat-kernel expansion. This corresponds to a partial re-summation that leaves, however, terms proportional to powers of the ground state in higher order coefficients. For our calculation, this ``level'' of re-summation is sufficient. For other purposes (e.g., if one wishes to consider higher dimensionalities) it may be necessary to go beyond this. In the following we shall show how this procedure can be carried out systematically and here present a slightly more general expression than (\ref{exphk}). We start from the following expansion 
\bea
\mathsf{K}\left(z|x,x,\right) 
=
{g^{1/2}(x)  \over (4\pi)^{d/2}} {\Gamma(d/4) \over 2 \Gamma(d/2)} \sum_{m=0}^\infty z^{m-d/2 \over M} 
e^{- \sqrt{z} \mathsf{T}_{1/2} - z \mathsf{T}_1}
\hat{\mathsf{E}}_{2m}(x),
\label{newhkexp}
\eea
where we have inserted the exponential $\exp\left\{- \sqrt{z} \mathsf{T}_{1/2} - z \mathsf{T}_1\right\}$. Notice the additional term $\propto \mathsf{T}_{1}$. Expanding the exponential in (\ref{newhkexp}) and comparing with (\ref{hkexp}) we can extract the first few coefficients $\hat{\mathsf{E}}_{2m}(x)$. If we require that powers of $\mathsf{D}$ vanish in the new coefficients $\hat{\mathsf{E}}_{2m}(x)$,
\bea
\hat{\mathsf{E}}_{0}(x) &=& 1,~~~ \hat{\mathsf{E}}_{2}(x) = 0 \\
\hat{\mathsf{E}}_{4}(x) &=& {1\over (4\pi)^{d/2}}{\Gamma(d/4) \over 4 \Gamma(d/2)}
\left\{
(d-2)\left(
{\hat{\mathsf{1}}\over 90} R_{abcd}^2 - {\hat{\mathsf{1}}\over 90} R_{ab}^2
+ {\hat{\mathsf{1}}\over 6} \mathcal{R}_{ab}^2
+{\hat{\mathsf{1}}\over 15} \Delta R\right)
+{{1}\over 6} \mathsf{D}^{ab} R_{ab}
 \right.\\ \nonumber
&&\left.
{1\over d+2}\left(
{1\over 2}\mathsf{D}_{ab}\mathsf{D}^{ab} + {2\over 3}(d+1) \nabla_a\nabla_b\mathsf{D}^{ab} + {1\over 6}(d+4)\Delta \mathsf{D} 
\right) -2 \mathsf{P}
\right\}. 
\eea
leads to
\bea
\mathsf{T}_{1/2} &=& - {2 \Gamma(d/4+1/2) \over \Gamma(d/4)} \left\{{1\over 2d}\mathsf{D} +{\hat{\mathsf{1}}\over 6}R\right\}, \label{thalf}\\
\mathsf{T}_{1} &=& 
{1\over 32} \left( - {4\over 2+d} + {\Gamma^2\left({2+d\over 4}\right) \over \Gamma^2\left(1+ {d\over 4}\right)} \right) \mathsf{D}^2 
+
{1\over 72} \left( 2 - d + 4 {\Gamma^2\left({2+d\over 4}\right) \over \Gamma^2\left({d\over 4}\right) } \right) R^2 
+
{1\over 48} \left( -4 +
d {\Gamma^2\left({2+d\over 4}\right) \over \Gamma^2\left(1+ {d\over 4}\right)}
\right) \mathsf{D} R. ~~~~~~~~~~~~~~~
\label{tone}
\eea
One thing that should be noted is that derivatives appear starting from the coefficient $\hat{\mathsf{E}}_{4}$ which has mass dimension dim$\left[\hat{\mathsf{E}}_{4}\right] = 4$; wihtin our assumption of smooth ground state (in other words, smooth enough background space), derivatives appear as corrections at order 4 in the asymptotic expansion. We are not dealing with this aspect here, but such corrections can be included systematically by including the derivatives we have dropped in the computation of the determinant and by adding progressively more terms. Up to the order we have considered here, there is no change due to derivatives.

Finally, let us stress that while we are not claiming that the choice (\ref{newhkexp}), (\ref{thalf}), (\ref{tone}) fully re-sums powers of $\mathsf{D}$ at all orders, it does so up to $m=2$. We also notice, as anticipated, that the exponentiation has simplified some of the coefficients, yielding $\hat{\mathsf{E}}_{2}(x) = 0$ (to achieve this it is sufficient to insert the term $\propto \mathsf{T}_{1/2}$ alone). While it may be interesting to attempt at performing a full re-summation by adjusting the exponentiation, here we take a more modest point of view, and use the modified expansion (\ref{exphk}) only as a tool to extract certain non-analytic terms in the effective action.

\subsection{Zero- and finite-temperature integrals}
Using formulae (13), (14) and (15) in the main text, we can separate out the zero-temperature from the finite-temperature contributions to the zeta function:
\bea
\zeta_{\mathsf{D}_\pm}^{(1)}(s) &=& \left({\beta/(2m) \over (4\pi)^{(d+1)\over 2}}{\Gamma(d/4) \over \Gamma(d/2)}\right) \sum_{q=0}^\infty \left[
{1\over 2\Gamma(s)}\int_0^\infty {dz} 
 z^{{q\over 2}-{d\over 4}-{3\over 2}+s} \int {dv} 
e^{-\sqrt{z} \mathsf{T}_\pm}
\hat{\mathsf{E}}_{2q}\right]\nonumber\\
&& \equiv
\left({\beta /(2m)\over (4\pi)^{(d+1)\over 2}}{\Gamma(d/4) \over \Gamma(d/2)}\right)
\int {dv} \sum_{q=0}^\infty \mathscr{Z}^{(\pm)}_q (s)
\label{defZ}
\eea
for the zero-temperature part and
\bea
\zeta_{\mathsf{D}_\pm}^{(2)}(s) &=& 
\left({\beta/(2m) \over (4\pi)^{(d+1)\over 2}}{\Gamma(d/4) \over \Gamma(d/2)} 
\right) \sum_{q=0}^\infty \left[ 
{1\over \Gamma(s)}\int_0^\infty {dz} 
\sum_{n=1}^\infty \exp\left(- {\tilde\beta^2 n^2\over 4 z}\right)
 z^{{q\over 2}-{d\over 4}-{3\over 2}+s} \int {dv} 
e^{-\sqrt{z} \mathsf{T}_\pm}
\hat{\mathsf{E}}_{2q}\right] \nonumber\\
&&\equiv
\left({\beta /(2m)\over (4\pi)^{(d+1)\over 2}}{\Gamma(d/4) \over \Gamma(d/2)}\right)
\int {dv}\sum_{q=0}^\infty \mathscr{T}^{(\pm)}_q (s) 
\label{defT}
\eea
for the finite-temperature one. The above relations define the integrals $\mathscr{Z}^{(\pm)}_q (s)$ and $\mathscr{T}^{(\pm)}_q (s)$. Also, notice that since we are using zeta-function regularization, in (\ref{defZ}) we are implicitly assuming that $\Re s$ to be sufficiently negative - this allows to commute summations and integration prior to analytical continuation \cite{Elizalde:1994gf}. Using (\ref{defZ}) and (\ref{defT}) it is straightforward to arrive at formula (16) of the main text, i.e.
\bea
\zeta_{\mathsf{D}_\pm}(s) =
{ {\beta / (2m)}\over  \left({4 \pi}\right)^{1+d\over 2}}
 {\Gamma\left(d/4\right) \over \Gamma\left(d/2\right)}
\int {dv} \sum_{q=0}^\infty 
\left[
\mathscr{Z}^{(\pm)}_q (s) + \mathscr{T}^{(\pm)}_q (s)
\right]. \nonumber
\eea
The integrals can be calculated exactly:
 \bea
 \mathscr{Z}^{(\pm)}_q (s)
&=&
 \left[
{1\over 2\Gamma(s)}\int_0^\infty {dz} 
 z^{{q\over 2}-{d\over 4}-{3\over 2}+s} 
 e^{-\sqrt{z} \mathsf{T}_\pm}
\hat{\mathsf{E}}_{2q}\right] \nonumber\\
&=&
{\Gamma\left(2s + q -d/2 -1\right) \over \Gamma\left(s\right)} \left|\mathsf{T}_\pm\right|^{1-q+d/2-2s}\hat{\mathsf{E}}_{2q},~
\eea
where we have assumed $\Re\left[d-2(q+2s)\right] <-2$ prior to analytical continuation to $s\to 0$. As for the thermal contribution, we have
 \bea
 \mathscr{T}^{(\pm)}_q (s)
&=&
\left[ 
{1\over \Gamma(s)}\int_0^\infty {dz} 
\sum_{n=1}^\infty \exp\left(- {\tilde\beta^2 n^2\over 4 z}\right)
 z^{{q\over 2}-{d\over 4}-{3\over 2}+s}
e^{-\sqrt{z} \mathsf{T}_\pm}
\hat{\mathsf{E}}_{2q}\right] \nonumber\\
&=&\sum_{n=1}^\infty
\left\{
{\sigma_n^{2s+q-1-d/2}\over \Gamma(s)}
\left\{
\Gamma\left({2+d-2q-4s \over 4} \right)
{}_{0}\mathsf{F}_2 \left[ \left\{\right\}; \left\{{1\over 2},{1\over 2}-{d\over 4}+{q\over 2}+s\right\},-
{\mathsf{T}_\pm^2 \sigma_n^2 \over 4} \right] \right.\right. \nonumber\\
&&\left.- \mathsf{T}_\pm \sigma_n \; \Gamma\left( {d-2q-4s \over 4} \right)
{}_{0}\mathsf{F}_2 \left[ \left\{\right\}; \left\{{3\over 2},{1}-{d\over 4}+{q\over 2}+s\right\},-
{\mathsf{T}_\pm^2 \sigma_n^2 \over 4} \right]
\right\} \nonumber\\
&&
\left.+ {2 \Gamma\left(q+2s -1 -{d\over 2} \right)\over \Gamma(s)}  
\mathsf{T}_\pm^{1-q-2s+d/2}
{}_{0}\mathsf{F}_2 \left[ \left\{\right\}; \left\{1+{d\over 4}-{q\over 2}-s,{3\over 2}+ {d\over 4}-{q\over 2}-s\right\},-
{\mathsf{T}_\pm^2 \sigma_n^2 \over 4} \right] \right\}\hat{\mathsf{E}}_{2q},~~~~~~~~~~~~~
\eea
where the functions ${}_{0}\mathsf{F}_2 \left[ \left\{\right\}; \left\{x,y\right\}, z \right]$ are the generalized hypergeometric functions \cite{NIST:2023} and we have defined for brevity of notation $\sigma_n = n \tilde\beta /4$. The above expression contains the thermal corrections to the effective action. Here we won't be concerned with such corrections; but it is possible to show by Taylor expanding the above contribution and its first derivative for large $\tilde\beta$ that such contributions vanish in the zero-temperature limit.


\end{widetext}

\end{document}